\documentclass[numberedappendix,apj,twocolumn]{emulateapj}

\usepackage[colorlinks=true, citecolor=blue, linkcolor=black,breaklinks]{hyperref}

\newcommand{\lya}{Ly$\alpha$}
\newcommand{\ewlya}{$\mathrm{EW}_{Lya}$}
\newcommand\emissionline[2]{#1$\;${\scshape{#2}}}
\newcommand{\oiii}{[\emissionline{O}{iii}]}
\newcommand{\oii}{[\emissionline{O}{ii}]}

\newcommand{\obj}{MACS1149-JD}
\newcommand{\MJ}{MACS1149}

\newcommand{\gn}{GN-z11}
\newcommand{\os}{O16}

\newcommand{\msun}{$M_{\odot}$}
\newcommand{\HST}{\emph{HST}}
\newcommand{\spitzer}{\emph{Spitzer}}
\newcommand{\pz}{$P(z)$}

\newcommand{\zgrismpeaktext}{$z=9.53$} 
\newcommand{\zgrismtext}{$z_{\mathrm{grism}}=9.53^{+0.39}_{-0.60}$}
\newcommand{\zphot}{$9.51^{+0.06}_{-0.12}$}
\newcommand{\zphottext}{$z_{\mathrm{phot}}=9.51^{+0.06}_{-0.12}$}
\newcommand{\zcombined}{$9.49^{+0.06}_{-0.08}$} 
\newcommand{\zcombinedtext}{$z_{\mathrm{comb}} = 9.49^{+0.06}_{-0.08}$}

\newcommand{\medmu}{$18.6^{+4.3}_{-2.6}$} 
\newcommand{\medmuallmodels}{$9.0^{+41.0}_{-5.1}$} 
\newcommand{\medMUV}{$-18.5 \pm 0.1$} 
\newcommand{\medLUV}{$0.14 \pm 0.01$} 

\newcommand{\medsmass}{$5.8^{+0.9}_{-0.9}$} 

\newcommand{\medSFR}{$0.7^{+0.6}_{-0.2}$} 

\newcommand{\medAGE}{$340^{+29}_{-35}$} 

\newcommand{\lyafluxlimphotozglassI}{$5.3\times10^{-18}~\mathrm{erg\,s^{-1}cm^{-2}}$} 

\newcommand{\lyafluxlimphotozbest}{$1.8\times10^{-18}~\mathrm{erg\,s^{-1}cm^{-2}}$} 

\newcommand{\gpaI}{\mbox{P.A. $=32^{\circ}$}}
\newcommand{\gpaII}{\mbox{P.A. $=125^{\circ}$}}
\newcommand{\refsdalpaI}{\mbox{P.A. $=111^{\circ}$}}
\newcommand{\refsdalpaII}{\mbox{P.A. $=119^{\circ}$}}

\begin{document}

\title{HST Grism Observations of a Gravitationally Lensed Redshift 9.5 Galaxy}

\author{A. Hoag\altaffilmark{1},
M. Brada\v{c}\altaffilmark{1}, 
G. Brammer\altaffilmark{2},
K.-H. Huang\altaffilmark{1},
T. Treu\altaffilmark{3},
C.A. Mason\altaffilmark{3},
M. Castellano\altaffilmark{4},
M. Di Criscienzo\altaffilmark{4},
T. Jones\altaffilmark{1},
P. Kelly\altaffilmark{5},
L. Pentericci\altaffilmark{4},
R. Ryan\altaffilmark{2}
K. Schmidt\altaffilmark{6},
M. Trenti\altaffilmark{7}}

\altaffiltext{1}{Department of Physics, University of California, Davis, 1 Shields Ave, Davis, CA 95616, USA; \email{athoag@ucdavis.edu}}
\altaffiltext{2}{Space Telescope Science Institute, 3700 San Martin Drive, Baltimore, MD 21218, USA}
\altaffiltext{3}{Department of Physics and Astronomy, University of California, Los Angeles, CA 90095-1547, USA}
\altaffiltext{4}{INAF, Osservatorio Astronomico di Roma, via Frascati 33, I-00040 Monteporzio, Italy}
\altaffiltext{5}{Department of Astronomy, University of California, Berkeley, CA 94720-3411, USA}
\altaffiltext{6}{Leibniz-Institut f\"{u}r Astrophysik Potsdam (AIP), An der Sternwarte 16, 14482 Potsdam, Germany}
\altaffiltext{7}{School of Physics, University of Melbourne, Parkville 3010, VIC, Australia}

\begin{abstract}
We present deep spectroscopic observations of a Lyman-break galaxy candidate (hereafter MACS1149-JD) at $z\sim9.5$ with the \emph{Hubble} Space Telescope (\emph{HST}) WFC3/IR grisms. The grism observations were taken at 4 distinct position angles, totaling 34 orbits with the G141 grism, although only 19 of the orbits are relatively uncontaminated along the trace of MACS1149-JD. We fit a 3-parameter ($z$, F160W mag, and \lya{} equivalent width) Lyman-break galaxy template to the three least contaminated grism position angles using an MCMC approach. The grism data alone are best fit with a redshift of \zgrismtext{} (68\% confidence), in good agreement with our photometric estimate of \zphottext{} (68\% confidence). Our analysis rules out Lyman-alpha emission from MACS1149-JD above a $3\sigma$ equivalent width of 21 \AA{}, consistent with a highly neutral IGM. We explore a scenario where the red \emph{Spitzer}/IRAC $[3.6] - [4.5]$ color of the galaxy previously pointed out in the literature is due to strong rest-frame optical emission lines from a very young stellar population rather than a 4000 \AA{} break. We find that while this can provide an explanation for the observed IRAC color, it requires a lower redshift ($z\lesssim9.1$), which is less preferred by the \HST{} imaging data. The grism data are consistent with both scenarios, indicating that the red IRAC color can still be explained by a 4000 \AA{} break, characteristic of a relatively evolved stellar population. In this interpretation, the photometry indicate that a \medAGE{} Myr stellar population is already present in this galaxy only $\sim500~\mathrm{Myr}$ after the Big Bang.

\end{abstract}


\vspace*{0.4truecm}

\section{Introduction}

A current frontier in cosmology is understanding the formation and growth of the first stars and galaxies, which formed within the first billion years of the universe \cite[for a recent review, see][]{Stark16R}. With recent observations of thousands of galaxies at $z\gtrsim6$, models of galaxy formation in the early universe can now be directly tested. Observations of such galaxies have found evidence for established stellar populations in some cases \cite[e.g.][]{Egami+05,Richard+12,Zheng+12}, where the ages of these galaxies would imply that they formed as early as $<300~\mathrm{Myr}$ after the Big Bang. It is important to note that the ages of stellar populations at $z>6$ are inferred from rest-frame optical (observed mid-IR) photometry, which could be affected by strong nebular line emission \citep[e.g.][]{Schaerer+09,Shim+11,Labbe+13,Smit+14,Holden+16,Castellano+17}. The recent spectroscopic discovery of a very luminous and massive ($\sim10^9~$\msun{}) galaxy (hereafter \gn{}) at $z=11.09$ shows that such a galaxy can form only $\sim400~\mathrm{Myr}$ after the Big Bang \cite[][hereafter \os{}]{Oesch+16}, although this galaxy is likely younger than $100~\mathrm{Myr}$. 

While surveys have found hundreds of galaxies up to $z\sim8$ \citep{Bouwens+15}, \gn{} is one of only three galaxies with spectroscopically confirmed redshifts at $z>8$ \citep{Zitrin+15b,Laporte+17}. Spectroscopic confirmation is critical for interpreting the results because at such high redshift, contamination from lower redshift galaxies and other interlopers is a major concern. Spectroscopy of such galaxies has primarily been limited to searching for the Lyman-alpha line (\lya{}) or other weaker rest-frame ultraviolet (UV) emission lines \citep[e.g.][]{Stark+15,Zitrin+15c}. The \lya{} optical depth increases rapidly with redshift, predominantly due to absorption by the increasing fraction of neutral hydrogen \citep{Mesinger+15}, such that at $z\gtrsim8$, the probability of detecting \lya{} emission in a single galaxy is very low \cite[e.g.][]{Pentericci+11,Pentericci+14,Schenker+12,Treu+13}. On one hand, the change in optical depth provides an important tool for constraining the end of cosmic reionziation. However, it also means that \lya{} emission is an inefficient tool for spectroscopic follow-up at $z>8$. 

The most distant confirmed galaxy, \gn{}, does not show \lya{} emission. Instead, \os{} detected a break in the spectral continuum using the \emph{Hubble} Space Telescope (\HST{}) G141 grism. They concluded that the break was most likely the \lya{} break, the strong absorption of flux blue-ward of \lya{} by neutral hydrogen in the inter-galactic medium (IGM). The technique of inferring the redshift from the location of the continuum break, while it does not suffer the same redshift-dependent effects as \lya{}, has its own challenges. Most importantly, one must securely detect the spectroscopic continuum red-ward of the break. For an $L_{\star}$ source at $z\sim8$, the apparent J-band AB magnitude is $m_{AB}\sim26.5$, clearly making this technique challenging for apparently faint sources. 

Gravitational lensing can magnify the more abundant intrinsically fainter sources and make it easier to detect their \lya{} break. In this work, we aim to spectroscopically confirm the redshift of a $m_{AB} = 25.7$ $z\sim9.5$ candidate \obj{}, originally discovered by \cite{Zheng+12}. The galaxy is magnified by a factor of $\mu =$ \medmu{} by the $z=0.545$ strong lensing cluster MACSJ1149.5+222 \citep[hereafter \MJ{};][]{Ebeling+01}. Since the discovery of \obj{}, \MJ{} has been observed extensively with \HST{} both in imaging and grism spectroscopy modes. The combination of extreme depth of the \HST{} grism data and the magnification of the source offer an unprecedented opportunity to study the rest-frame UV continuum of a sub-$L_{\star}$ galaxy at $z\sim9.5$. 

In Section~\ref{sec:data}, we present all of the imaging and spectroscopic data available for \obj{}. We describe the global contamination model we built from the grism data in Section~\ref{sec:contam}. We show our spectral extraction for \obj{} in Section~\ref{sec:extract} and the inferences on the redshift from the grism and photometry in Section~\ref{sec:redshift}. In Section~\ref{sec:discussion}, we discuss the implications of our results, and we conclude in Section~\ref{sec:summary}. Throughout this work, we adopt the following cosmology: $\Omega_m=0.3$,
$\Omega_{\Lambda}=0.7$ and $h=0.7$. All magnitudes are given in the AB system.

\section{Data}
\label{sec:data}
Here we describe the imaging and spectroscopic data that we used to study \obj{}. 

\subsection{Imaging data}

The filters and magnitudes used for photometry of \obj{} are listed in Table~\ref{tab:photometry}. These data came from the ASTRODEEP photometric catalog\footnote{\url{http://www.astrodeep.eu/ff34/}} \citep{astrodeep+17}, which used optical and near infrared \HST{} images from the Hubble Frontier Fields Initiative \citep[HFF;][]{Lotz+16}, Keck/MOSFIRE $K_s$ images from the K-band Imaging of the Frontier Fields (KIFF) program \citep{Brammer+16}, and \spitzer{}/IRAC $[3.6]$ (CH1) and $[4.5]$ (CH2) images from the \emph{Spitzer} Ultra-Faint SUrvey Program \citep[SURFSUP][]{Bradac+14} and from Director's Discretionary Time (PI Capak). The ASTRODEEP team performs subtraction of the intra-cluster light and cluster members to obtain more accurate photometry and improved number counts of lensed, high redshift sources. We refer to \citet{Merlin+16} and \citet{Castellano+16b} for further details on the photometry.


\begin{figure}[htb]
    \centering
    \includegraphics[width=\linewidth]{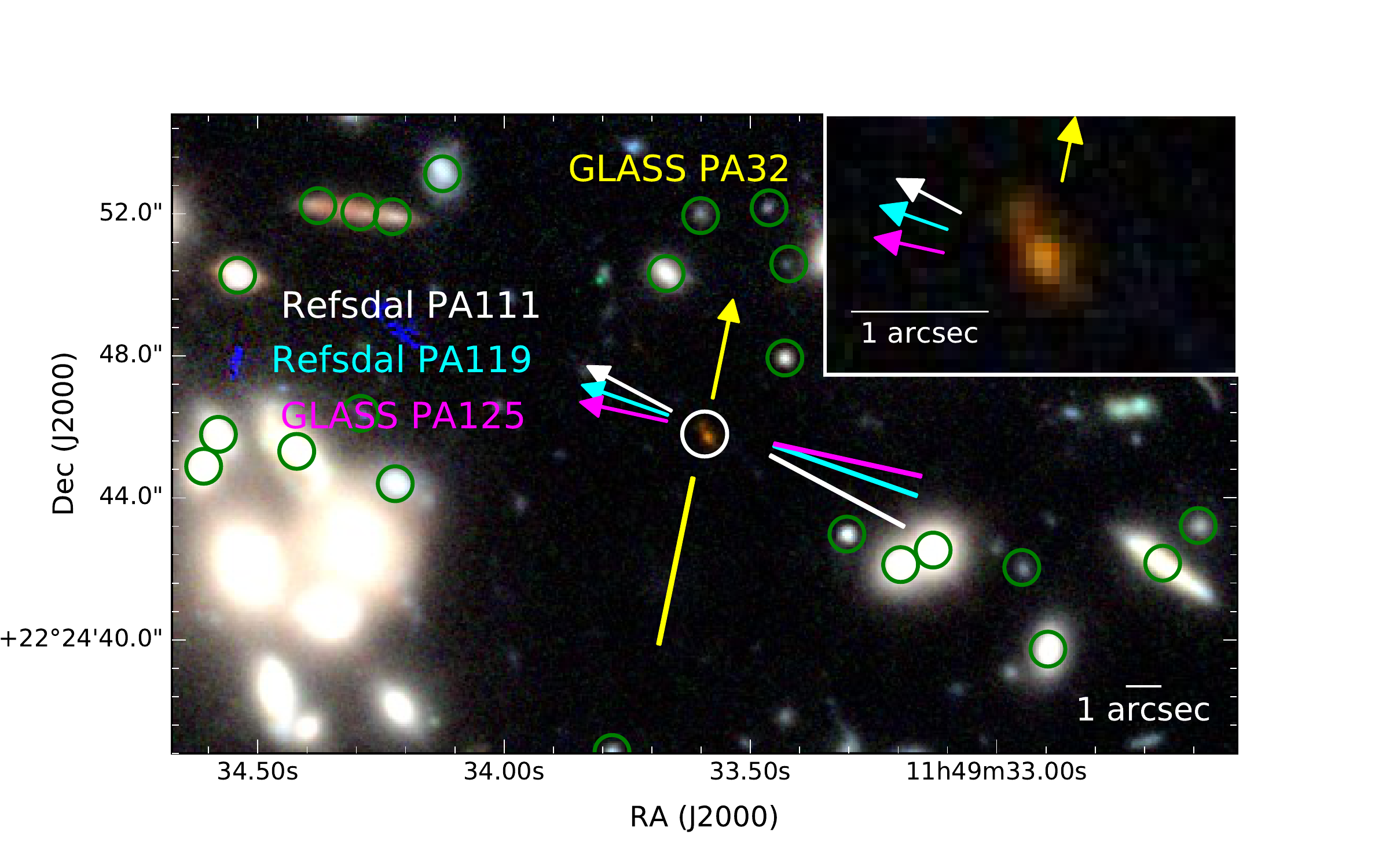}
    \caption{Color composite image from the \HST{} filters: F105W, F125W and F160W. \obj{} is circled in white in the center of the image. The 4 grism dispersion directions are shown with the colored arrows. The labels on the arrows correspond to the PA\_V3 header keyword, following the GLASS convention \citep{Treu+15a}. The cluster core containing the brightest cluster galaxy (not shown) is approximately $1'$ to the south of \obj{}. The green circles show a subset of the objects whose spectra (of any order) have the potential to contaminate the spectrum of \obj{}. We modeled these galaxies separately to obtain a global contamination model for \obj{}. A zoomed in region centered on \obj{} is shown in the upper right inset.}
 \label{fig:HST}
 \end{figure}
 

\begin{deluxetable}{ll}
\tablecaption{Photometry of \obj{} }
\tablecolumns{2}
\tablewidth{0.8\linewidth}

\tablehead{Filter & AB magnitude}

\startdata
F435W & $> 27.84$ \\ 
F606W & $> 28.38$ \\ 
F814W & $> 28.53$ \\ 
F105W & $> 28.67$ \\ 
F125W & $26.85 \pm 0.07$ \\ 
F140W & $26.00 \pm 0.05$ \\ 
F160W & $25.71 \pm 0.03$ \\ 
HAWK-I Ks & $> 25.06$ \\ 
IRAC CH1 & $> 25.52$ \\ 
IRAC CH2 & $24.79 \pm 0.15$ 
\enddata

\label{tab:photometry}
\tablecomments{ASTRODEEP \citep{astrodeep+17} photometry for \obj{} used in this work. Limiting magnitudes are $3\sigma$. \obj{} corresponds to source 2876 with $z_{\mathrm{phot}} = 9.56$ in the ASTRODEEP catalog of \MJ{}. }
\end{deluxetable}


\subsection{Slitless Grism Data}
In addition to the significant investment of \HST{} imaging, \MJ{} has been observed extensively with the \HST{} near-infrared (NIR) grisms. Table~\ref{tab:spectroscopy} provides an overview of the spectroscopy we use in this work. \MJ{} was first observed with the \HST{} G102 and G141 NIR grisms by the Grism Lens-Amplified Survey from Space \citep[GLASS;][]{Schmidt+14,Treu+15a} in 2014. GLASS obtained 10 orbits in G102 and 4 orbits in G141, which together provide approximately uniform continuum sensitivity over the wavelength range $0.81-1.69 \mu m$. The GLASS data were taken at two approximately orthogonal dispersion position angles to aid contamination subtraction and to provide two independent detections of real spectral features \citep{Treu+15a}. While \obj{} was not taken into consideration when the GLASS position angles were chosen, its first order spectra are contained in both with approximately equal exposure time in each position angle. We refer to the GLASS positions angles as \gpaI{} and \gpaII{} throughout\footnote{The P.A. values are taken from the PA\_V3 header keyword in the individual flt files, following the GLASS convention \citep{Treu+15a}. These differ from the actual dispersion position angles measured E from N ($\mathrm{PA}_\mathrm{disp}$). One can easily be obtained from the other by $\mathrm{PA}_\mathrm{disp} = \mathrm{PA}_\mathrm{V3} - 45.2$.}. 

The discovery of Supernova (SN) Refsdal \citep{Kelly+15} in the \MJ{} prime field prompted a \HST{} Director's Discretionary Time program to obtain 30 additional orbits in G141 during 2014 December - 2015 January to determine the SN type \citep[HST-GO-14041;][]{Kelly+16} (hereafter the ``Refsdal'' data). The Refsdal data were taken at two position angles (\refsdalpaI{}, \refsdalpaII{}), which were optimized to reduce the contamination for extracting the spectrum of SN Refsdal. Both of these position angles contain the spectra of \obj{} with approximately equal exposure time in each position angle. However, both position angles are separated by less than $25^{\circ}$ from the GLASS \gpaII{}, the GLASS position angle in which \obj{} is more contaminated. The dispersion directions corresponding to the two GLASS and two Refsdal position angles are indicated by the arrows in Figure~\ref{fig:HST}. 

The GLASS and Refsdal grism data were reduced together using the Grism redshift \& line analysis software for space-based slitless spectroscopy ({\tt Grizli}\footnote{\url{https://github.com/gbrammer/grizli/}}; Brammer et al. in preparation). We briefly outline the data reduction steps here. First, raw grism and direct exposures were obtained from the Mikulski Archive for Space Telescopes (MAST\footnote{\url{https://archive.stsci.edu/index.html}}). In addition to grism exposures, both GLASS and Refsdal programs obtained direct images in \HST{} to establish the wavelength zero points of the grism exposures. F105W and F140W were used to calibrate G102 and G141, respectively, due to similarity in wavelength coverage. The HFF F814W v1 mosaic was used for absolute astrometric calibration. Raw exposures containing bright satellite trails and high backgrounds due to the He Earth-glow were manually identified and discarded \citep{Brammer+14}. Variable sky backgrounds were accounted for as in \cite{Brammer+12}. For each individual exposure we mask pixels where the MAST persistence model\footnote{\url{https://archive.stsci.edu/prepds/persist/}} predicts persistent flux at a level greater than $0.6$ times the pixel flux uncertainty \citep{Brammer16inst}.
 
Grism data from different visits were combined using Astrodrizzle \citep{Gonzaga+12}. The drizzled products consist of a two-dimensional cutout with two orthogonal axes: a spatial axis along the vertical and a calibrated wavelength axis along the horizontal. This process rectifies the trace of the spectrum along the horizontal axis, as opposed to, for example, the GLASS products and the spectra in \os{}, which were both reduced using the 3D-HST pipeline \citep{Brammer+12} and have a noticeable spectral tilt. The grism products have, before drizzling, a spectral dispersion of $24.5$ \AA{}/pix and $46.5$ \AA{}/pix in G102 and G141, respectively. The G102 and G141 data provide a continuous spectrum of \obj{} from $0.81-1.69 \mu m$, although with varying sensitivity as a function of wavelength because the combined G141 spectrum is significantly deeper than G102. Nevertheless, the G102 data are included in this analysis as they are helpful for building the contamination model. 

\begin{deluxetable}{lllll}
\tablecaption{Spectroscopic data used in this work }
\tablecolumns{5}
\tablewidth{\linewidth}

\tablehead{Grism & P.A. ($^{\circ}$) & $N_{\mathrm{orbits}}$ & $t_{\mathrm{exp}} (s)$ & Program}

\startdata
G102 & 32 & 5 & 9826 & GLASS\tablenotemark{$a$}  \\
G102 & 125 & 5 & 10629 & GLASS \\
G141 & 32 & 2 & 4412 & GLASS \\
G141 & 125 & 2 & 4412 & GLASS \\
G141 & 111 & 15 & 35788 & Refsdal\tablenotemark{$b$} \\
G141 & 119 & 15 & 34988 & Refsdal
\enddata

\tablecomments{The P.A. values are taken from the PA\_V3 header keyword in the individual flt files, following the GLASS convention. Exposure times are rounded to the nearest second. }
\tablenotetext{$a$}{The ``GLASS'' data refer to the Grism Lens-Amplified Survey from Space (HST-GO-13459; PI Treu).} 
\tablenotetext{$b$}{The ``Refsdal'' data refer to the Supernova Refsdal follow-up program (HST-GO-14041; PI Kelly).}

\label{tab:spectroscopy}

\end{deluxetable}


\section{Contamination Modeling}
\label{sec:contam}

We used {\tt Grizli} to build a global contamination model by modeling each source for each visit and position angle. By working at an early stage in the reduction, the noise properties of each model unit are characterized more readily than after drizzling, which can produce correlated noise. Models from each visit can then be drizzled together for extraction of spectral features from the full-depth data. 

The global contamination model requires three components: a reference image for image/grism alignment, a segmentation image and a Source Extractor \citep[SExtractor;][]{Bertin+96} photometric catalog associated with the segmentation image. For the reference image, we used a full depth \HST{} WFC3/IR F160W mosaic created by co-adding all of the available F160W for \MJ{} data on MAST, as of November 2016.  We used the segmentation map and associated source catalog provided by GLASS\footnote{Available here: \url{https://archive.stsci.edu/prepds/glass/}}. The segmentation map was created from the stack of the CLASH NIR bands using the CLASH SExtractor parameters, and the GLASS team used this segmentation map to perform contamination subtraction on their own data \citep{Schmidt+14,Treu+15a}. We did not notice a significant difference in the resulting model when using a segmentation image created from the full-depth HFF images, as expected because the contamination is dominated by bright sources. 

First, a simple flat (in $F_{\lambda}$) spectrum model was assigned to every object in the segmentation map satisfying $H_{160}<25$, where the normalization of the model was determined by the flux in the F160W reference image. Then, a second order polynomial was fit to each object in the flat model satisfying $16<H_{160}<24$. This last step is iterated until differences in the global contamination model become negligible, which in our case occurred after three total iterations. 

In an attempt to more carefully model the contamination affecting the trace of \obj{}, we performed another tier of modeling on top of what is described above. We searched the segmentation map for all sources that could possibly contaminate the first order spectrum of \obj{} in each of the four position angles. This amounted to selecting all objects at least as bright ($H_{160}<25.6$) as \obj{} that fall within $\pm1\farcs8$ of the expected trace of \obj{} in all four position angles. The size of the search beam was chosen so that off-center extended objects would also be flagged. 164 objects satisfied these criteria. We then visually inspected each of these objects in the HFF v1 images as well as the segmentation image. We found that 14 of these objects were likely spurious, either because they lacked a counter-part in the deep \HST{} images or because they were part of a diffraction spike. 

For the 150 remaining objects (some shown in Figure~\ref{fig:HST}), which we deemed were primary contaminants of \obj{}, we refined their models further by performing spectral energy distribution (SED) fitting of the grism data. {\tt Grizli} performs SED fitting by comparing two-dimensional grism models from, in our case, a minimal set of galaxy templates to two-dimensional cutouts of each object's first order spectrum.  Briefly, we considered four templates: 

\begin{itemize}
\item A very young, low metallicity Lyman-break galaxy (LBG) based on  Q2343-BX418 \citep{Erb+10}
\item An intermediate age SED with a moderate 4000 \AA{} break (D4000)
\item An old simple stellar population from \cite{Conroy+12}
\item A post-starburst template with prominent Balmer break and strong D4000
\end{itemize}

In addition to the templates, a set of emission lines with fixed line ratios are included in the fit. The emission lines are useful for breaking redshift degeneracies. The redshift fitting was first run over a coarse grid ($\Delta z = 0.01$) over the interval $0.5<z<2.3$. This range was chosen based on the reliability of detecting strong emission lines in G102 and G141 at these redshifts. The templates are flexible enough to provide an improved description of the continua over a simple polynomial model even if the redshifts are incorrect. A second, finer grid ($\Delta z = 0.002$) was run  
around the best-fit peaks (determined by the $\chi^2$) in the first grid. The best-fit redshift found during the second, finer redshift grid was assigned to the galaxy. 

\begin{figure}[]
    \centering
    \includegraphics[width=\linewidth]{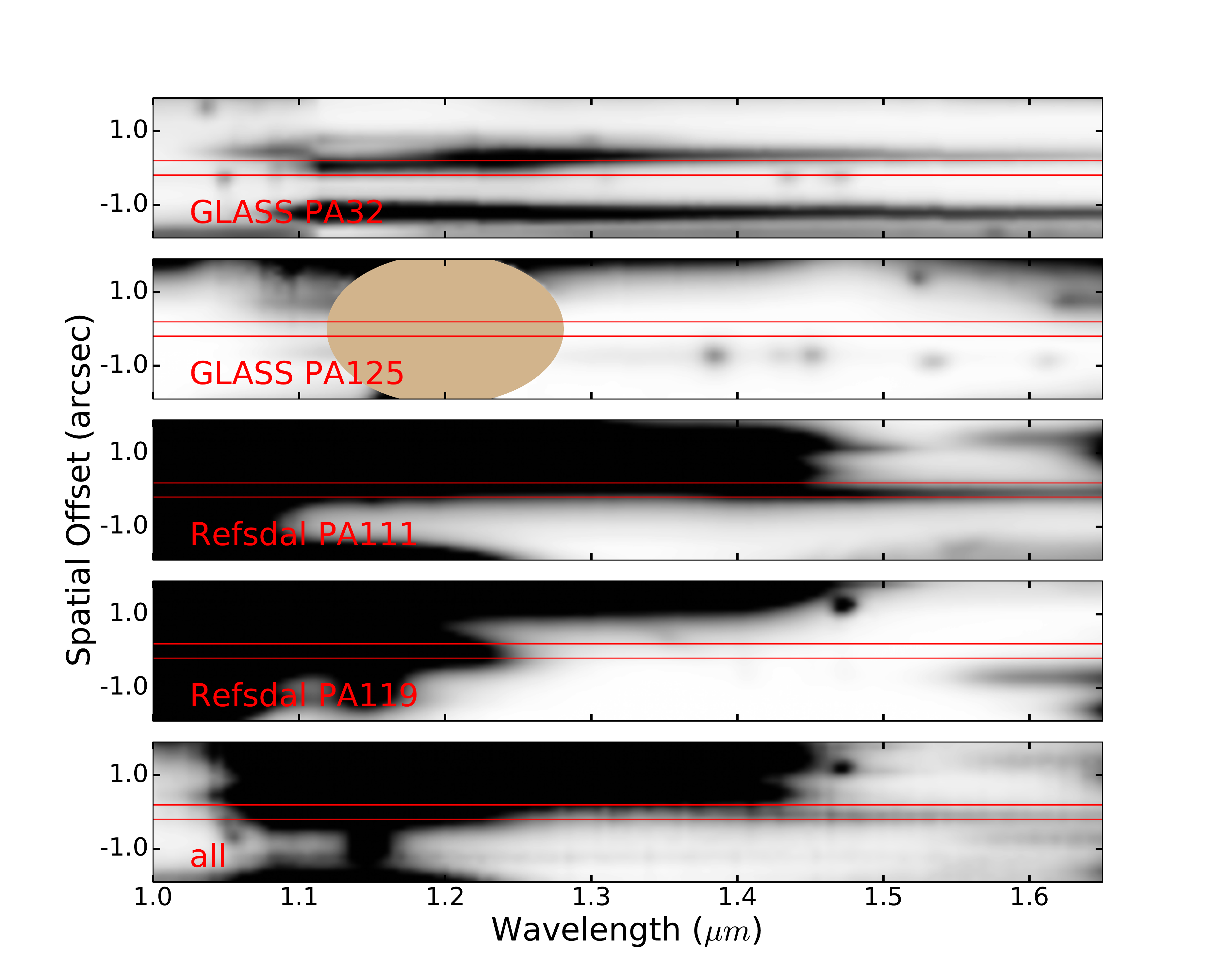}
    \caption{Contamination models for each of the 4 individual orients as well as the stack (bottom) extracted near the trace of \obj{} (red horizontal lines). \refsdalpaI{} is severely contaminated over the entire trace. The contamination is dominated by the second order trace of a bright object, making the contamination in this orient especially difficult to model. Black represents positive signal. The tan ellipse in GLASS \gpaII{} masks out a large detector artifact.}
 \label{fig:contam}
 \end{figure}

The top four panels of Figure~\ref{fig:contam} show the drizzled contamination models from each position angle at the location of the first order spectrum of \obj{}. The bottom panel shows the stack of all four position angles. Although the two Refsdal position angles are 15 orbits each in G141, compared to the 2 orbits each for the two GLASS position angles, the Refsdal \refsdalpaI{} is severely contaminated along the trace of \obj{}. Furthermore, the contamination in this P.A. is dominated by the second order spectrum of a bright galaxy (not shown in Figure~\ref{fig:HST}). {\tt Grizli} fits to the first order spectra, while the second order spectra of contaminants are determined from a calibration between first and higher order spectra, which is in general less accurate than direct modeling. The other three position angles are much cleaner of contamination. As a result, we discard the Refsdal \refsdalpaI{} data in our analysis. 


\section{Continuum Extraction}
\label{sec:extract}

\begin{figure}[htb]
    \centering
    \includegraphics[width=\linewidth]{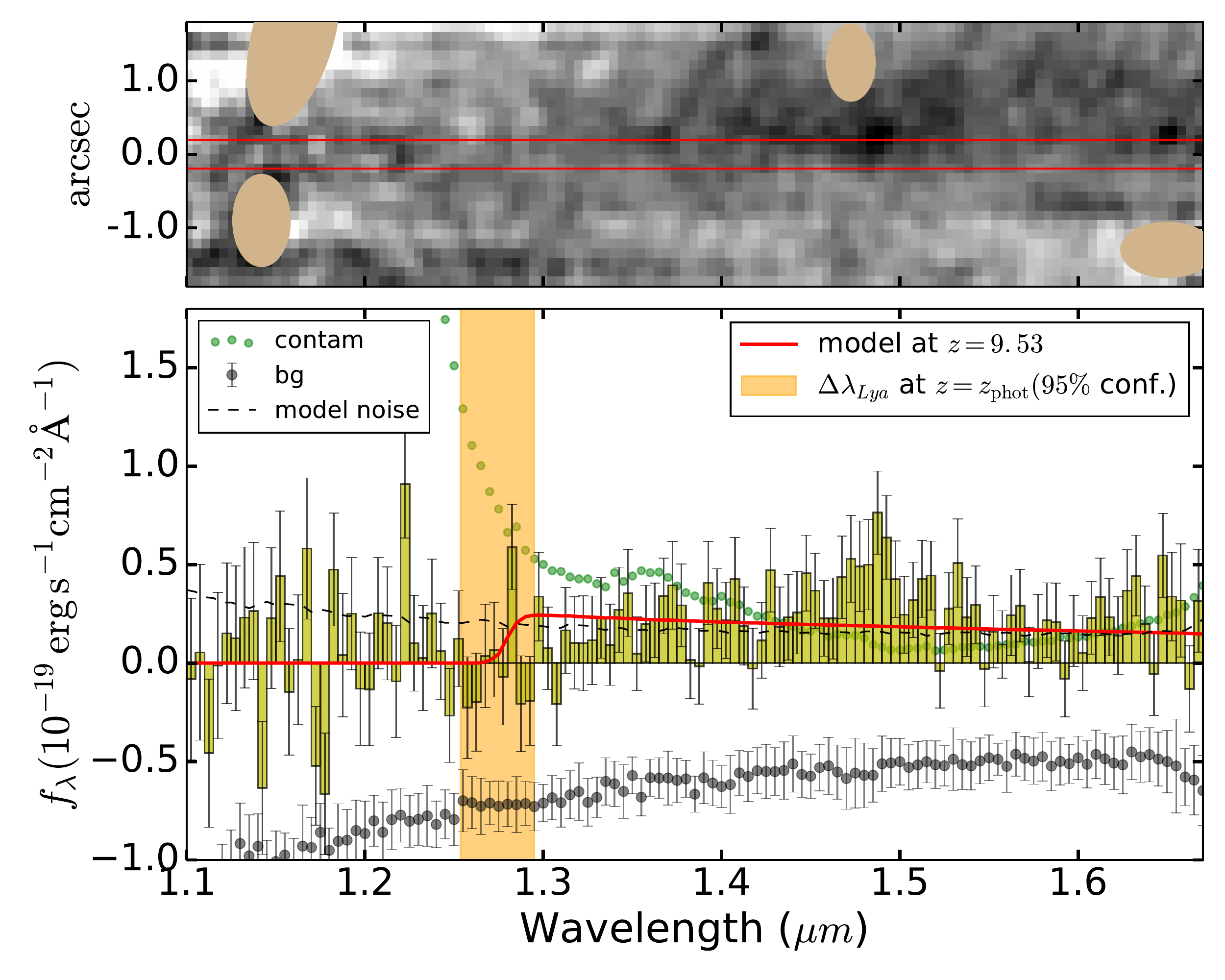}
    \caption{\textbf{Top}: 2D contamination-subtracted grism spectrum of \obj{} from the 3 reliable position angles. Black represents positive signal. The tan ellipses mask out 0th order contamination residuals. The 2D spectrum was smoothed with a 2D Gaussian kernel with $\mathrm{FWHM}=2$ pixels in both dimensions to bring out the continuum feature. \textbf{Bottom}: 1D spectrum (yellow histogram) extracted from the spatial aperture denoted by the red lines in the top panel. The red line shows the predicted signal from a Lyman-break galaxy model at the most probable redshift derived from the grism data (\zgrismpeaktext{}, see Section~\ref{sec:mcmc}), normalized to the \HST{} F160W magnitude. The black dashed line shows the predicted noise level in the absence of contamination. The shaded yellow region indicates the wavelength range where \lya{} would be expected from the 95\% confidence interval on the photometric redshift. While we do not detect \lya{} emission, a positive signal is detected above the noise level at $>1.25~\mu m$, which is likely the continuum from \obj{}. Both the 2D and 1D spectra are drizzled to $50$ \AA{} $/$ pixel. }
 \label{fig:spec_extract}
 \end{figure}

We show the two-dimensional contamination-subtracted spectrum of \obj{} combined from the two GLASS position angles and the single Refsdal \refsdalpaII{} in the top panel of Figure~\ref{fig:spec_extract}. The final spectrum is an inverse variance weighted stack of the 3 position angles, totaling 19 orbits of G141. The contamination was not included in the variance array when stacking (see Section~\ref{sec:mcmc}). However, we  down-weighted an especially contaminated off-trace portion of the GLASS \gpaI{} spectrum for display purposes only (see the Appendix).

The one-dimensional spectrum shown in the bottom panel of Figure~\ref{fig:spec_extract} was obtained by summing within a 3 pixel ($0\farcs38$) aperture along the spatial (vertical) axis of the two-dimensional spectrum. A positive signal is detected in the one-dimensional spectrum at $\lambda \gtrsim1.25~\mu m$. The median signal-to-noise ratio ($S/N$) per $50$ \AA{} bin is $1.1$ over the continuous interval $1.25<\lambda<1.65~\mu m$, whereas it is $S/N = 0.2$ at $\lambda<1.25~\mu m$. The integrated $S/N$ over the interval $12500<\lambda<16500$, is $(S/N)_{\mathrm{total}} = 11.1$. We also performed a \citet{Horne+86} extraction, but the increase in $S/N$ of the detection was negligible in comparison to the simple sum. In Figure~\ref{fig:spec_extract} we also show the one-dimensional contamination spectrum and rms noise spectrum extracted using the same size spatial aperture. 

In the process of extracting the one-dimensional spectrum, we also subtracted a local background, which we computed by taking the median in each column of the data array after masking out contaminated pixels. We used a variety of thresholds when masking to assess the change in the background. We varied the threshold based on the fraction of the mean contamination level in the entire two-dimensional contamination array, ranging from 0.05 to 1 times the mean contamination, in steps of 0.05. We then applied these masks to the data array and calculated the median in each column. The median and standard deviation of the backgrounds calculated at each threshold level were used as the background and the error on the background in our spectral extraction. Before background subtraction, the mean value of the pixels outside of the trace of \obj{} was \mbox{$-1.51^{+1.08}_{-0.37}\, \times 10^{-19}~\mathrm{erg\,s^{-1}\,cm^{-2}\,pix^{-1}}$}, whereas after background subtraction we obtained \mbox{$-0.01^{+0.29}_{-0.19}\, \times 10^{-19}~\mathrm{erg\,s^{-1}\,cm^{-2}\,pix^{-1}}$}. Therefore, we successfully subtracted the negative background, although some slight positive off-trace residuals remain, which we discuss in the Appendix. We note that the error bars on the one-dimensional spectrum shown in Figure~\ref{fig:spec_extract} incorporate the rms noise and the uncertainty on the background. 

We show several tests of the validity of the continuum detection in the Appendix. Most importantly, when the GLASS and Refsdal data are separated, the bulk of the positive signal observed in Figure~\ref{fig:spec_extract} arises from the deeper Refsdal data (Figure~\ref{fig:glassvsrefsdal2}). While there are contamination residuals away from the trace of \obj{} (Figure~\ref{fig:offsets}), the lack on contamination along the trace of \obj{} indicates that the positive signal is unlikely to arise from a contamination residual.


\section{Redshift estimation}
\label{sec:redshift}

\subsection{Photometric Redshift}
\label{sec:photoz}

The photometric redshift was measured using the methodology described by \citet{Huang+16a}. In brief, we used the code EAZY \citep{Brammer+08} along with the stellar population templates from \citet[][hereafter~\textit{BC03}]{BC03}, a \citet{Chabrier03} initial mass function between $0.1$ and $100 M_{\odot}$, a \citet{Calzetti+00} dust attenuation law, and an exponentially declining star formation history (SFH) with a fixed metallicity of 0.2 $Z_{\odot}$. Nebular emission lines were added to the galaxy templates as in \citet{Huang+16a}. 

\begin{figure}[htb]
    \centering
    \includegraphics[width=\linewidth]{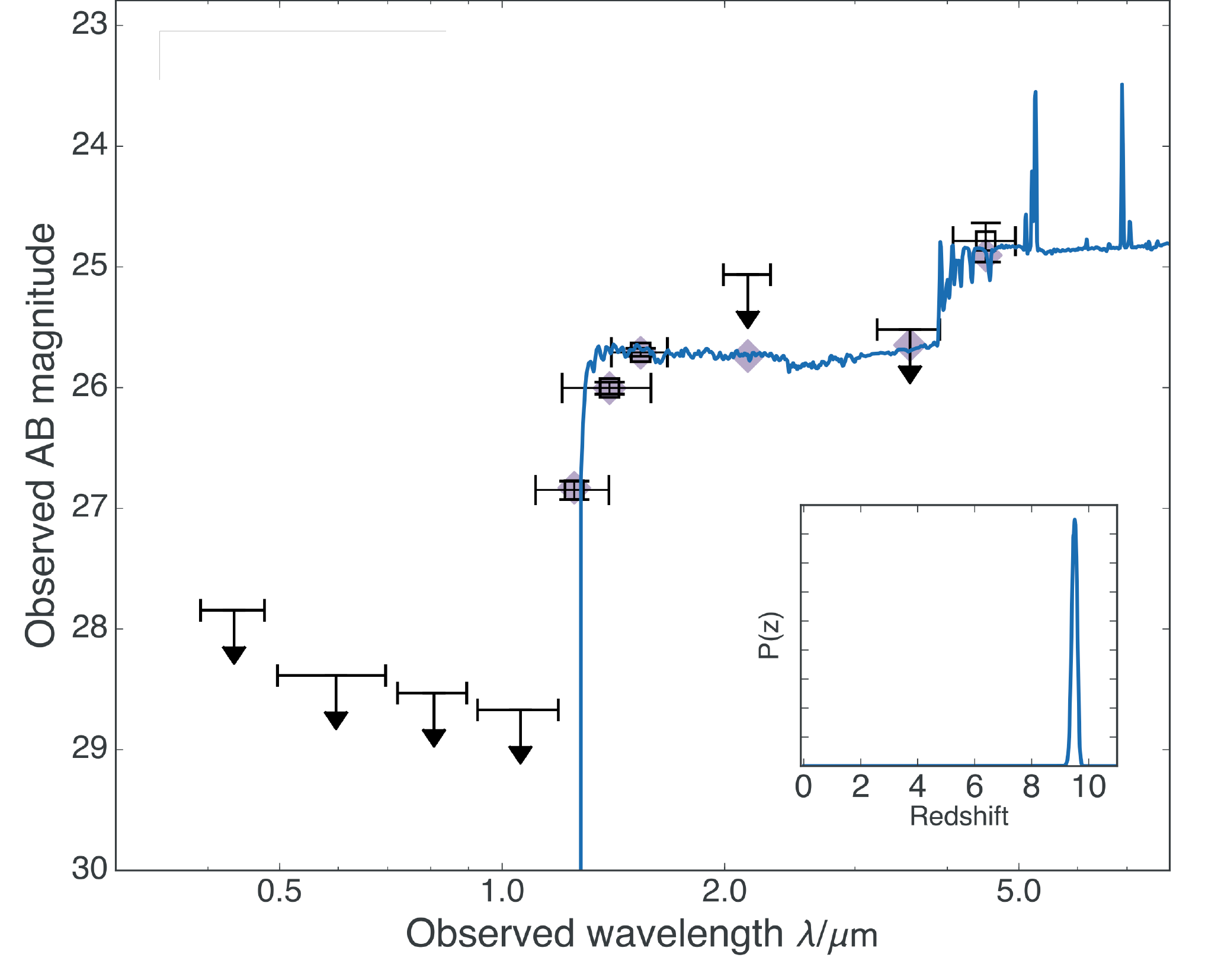}
    \caption{Best-fit spectral energy distribution (SED) for \obj{} from ASTRODEEP photometry. The purple diamonds represent the magnitudes of the SED integrated over each filter bandpass. The photometric data are summarized in Table~\ref{tab:photometry}. Downward facing arrows represent $3\sigma$ upper limits. }
 \label{fig:SED}
 \end{figure}

The photometric data used to constrain the redshift are from the ASTRODEEP catalogs and are summarized in Table~\ref{tab:photometry}. The data are extremely deep and span the interval from $\sim0.4-5 \mu m$.  Figure~\ref{fig:SED} shows the best-fit spectral energy distribution (SED) template and the photometric detections and $3\sigma$ upper limits for \obj{}. The photometric redshift distribution (hereafter \pz{}) is shown in the inset in Fig~\ref{fig:SED} below the SED. The best-fit photometric redshift is $z=$ \zphot{} (68\% confidence). This is in agreement with recent estimates of the photometric redshift, for example $z=9.44\pm0.12$ obtained by \citet{Zheng+17}, who used the same imaging data except for the MOSFIRE Ks-band, which does not provide strong additional constraints on the SED. The \spitzer{}/IRAC CH1 and CH2 data are much more important. For example, if we exclude CH1 and CH2 from our photometry, the 68\% confidence interval on the photo-$z$ is $[9.22,9.68]$, a factor of $2.5$ broader than when including CH1 and CH2. 

Interestingly, \citet{Zheng+17} report a $S/N=5.9$ detection in IRAC CH1, i.e. a magnitude of $25.64\pm0.17$. In the ASTRODEEP catalog (Table~\ref{tab:photometry}), \obj{} is not detected in CH1 above the $3\sigma$ level. We also performed independent photometry on the full depth HFF HST and IRAC data using the method described by \citep{Huang+16a}, similarly not detecting \obj{} in CH1 above $3\sigma$. Whether CH1 is considered a detection or not, this does not seem to significantly affect the inferred photometric redshift; the photo-$z$'s obtained by \citet{Zheng+17} and ASTRODEEP are in statistical agreement, as we saw above. \citet{Huang+16a} obtained a slightly lower photometric redshift, $z=9.3\pm0.1$ for \obj{}, but using shallower IRAC data than in \citet{Zheng+17} and in this work. 


\subsection{Search for Lyman-alpha emission}

We first searched the grism data for emission lines within the wavelength range inferred from the photometric redshift, $9.31<z<9.65$ (95\% confidence). This interval corresponds to the wavelength range $1.254<\lambda<1.295~\mu m$, which is denoted by the yellow shaded region in the bottom panel of Figure~\ref{fig:spec_extract}. The spectrum is consistent with the noise in this interval, so \lya{} is not significantly detected above our flux limit. 

To estimate the \lya{} flux limit, we used the same method and the same aperture ($0\farcs6 \times 100$ \AA{}) as \cite{Schmidt+16}, who calculated flux limits for 47 $z\gtrsim7$ LBG candidates in 6 of the GLASS clusters, MACS1149 not included, throughout G102 and G141. \cite{Schmidt+16} found median $1\sigma$ \lya{} flux limits of $4-6\times10^{-18}~\mathrm{erg\, s^{-1} cm^{-2}}$ (95\% confidence) per 2-orbit G141 position angle in the photo-$z$ wavelength range, $1.254<\lambda<1.295~\mu m$, before taking into account magnification. As a cross-check, we measured the $1\sigma$ flux limit in this wavelength range using the GLASS \gpaI{} orient alone, finding \lyafluxlimphotozglassI{}, in agreement with the range found by \cite{Schmidt+16} for observations of the same depth. 

With the addition of the Refsdal data, we achieve a more sensitive \lya{} flux limit of \lyafluxlimphotozbest{} ($1\sigma$), before taking into account magnification. The limit was obtained using the GLASS \gpaI{} and Refsdal \refsdalpaII{} data, but not the other two position angles. The Refsdal \refsdalpaI{} data were not included for the reasons described in Section~\ref{sec:extract} and in the Appendix, and the GLASS \gpaII{} data were not included because of a detector artifact present in the wavelength range where \lya{} would be expected from the photometric redshift (Figure~\ref{fig:contam}). The \lya{} flux limit can be converted into a limit on the \lya{} equivalent width (EW) since the continuum redward of the expected position of the \lya{} line and the \lya{} break is detected in \HST{} imaging. Within the wavelength range $1.254<\lambda<1.295~\mu m$, the only \HST{} filter sampling the rest-frame UV in which the \lya{} emission line and the \lya{} break do not fall is F160W. Adopting the F160W magnitude as the baseline continuum level, we estimate a $3\sigma$ rest-frame \lya{} EW limit of $21$ \AA{}. For comparison, the same limit is $63$ \AA{} in the GLASS \gpaI{} data alone. \cite{Bian+15} obtained ground-based spectroscopy with LBT/LUCI \citep{Ageorges+10,Hill+12} of \obj{} over the wavelength interval $1.17-1.30~\mu$m, and similarly did not detect \lya{} emission. For comparison, they obtained $3\sigma$ rest-frame \lya{} EW limits of $\sim30$ \AA{} over the interval $1.17 < \lambda < 1.26~\mu$m and $\sim50$ \AA{} over $1.26 < \lambda < 1.30~\mu$m.

We calculated the probability of observing this EW limit in the context of a likely highly neutral universe \citep{Planck+16,Greig+17} using the method from \citet{Mason+17}. In brief, we take reionization simulations of the IGM by \citet{Mesinger+16} for different values of the neutral hydrogen fraction, $f_{\mathrm{HI}}$, and populate halos with LBG properties calibrated using the observed \lya{} EW distribution at $z=6$ to obtain EW distributions as a function of $f_{\mathrm{HI}}$ and $M_{UV}$. We consider two scenarios: 1) a completely ionized universe, i.e. $f_{\mathrm{HI}} = 0$ and 2) a more realistic universe with $f_{\mathrm{HI}} = 0.75$. We find $p(\mathrm{EW}<7$ \AA{}$) = 0.32$ in case 1) and $p(\mathrm{EW}<7$ \AA{}$) = 0.52$ in case 2). Hence the non-detection is not surprising in either case, but especially so in the mostly neutral universe. We note that the simulations are for $z=7$. At $z=9.5$, the \lya{} line from this galaxy would be further damped since the decreased cosmic volume would put \obj{} in closer proximity to neutral patches. Thus at $z=9.5$ our observation is even more likely than the numbers quoted above for $z=7$.  
  
\subsection{MCMC analysis}
\label{sec:mcmc}

Because the grism data are sensitive enough to detect the continuum of \obj{}, the search for the \lya{} emission line in the remaining portion of the grism data is more complicated; a detection of \lya{} emission is degenerate with a marginal detection of the continuum. To quantify this degeneracy, we performed a Markov chain Monte Carlo (MCMC) analysis with redshift ($z_g$), rest-frame \lya{} EW (\ewlya{}), and F160W magnitude ($m160$) as parameters. 

We generated one-dimensional LBG models with fixed rest-frame UV slope of $\beta=-2$ \citep[c.f.][]{Bouwens+14} at $z=0$ and set all flux blue-ward of the rest-frame \lya{} wavelength to zero. We then redshifted these templates to $z_g$ and added a \lya{} emission line with a rest-frame EW of \ewlya{}. The intrinsic FWHM of the line we used is $150~\mathrm{km}~s^{-1}$, although due to the low resolution of the grism the FWHM does not significantly influence the results. We normalized the models so that their integrated F160W magnitude equaled $m160$. The one dimensional models were then convolved with the object's 2D morphology for each individual exposure (with its given P.A.) and the exposure models were drizzled exactly as with the data spectrum shown in Figure~\ref{fig:spec_extract}. While the object is clearly resolved, as is visually apparent in Figure~\ref{fig:HST}, the compact central clump dominates the surface brightness. Because the surface brightness distribution of the object is used when drizzling the 1D models to 2D, the extended portion of the galaxy does not significantly contribute to the 2D drizzled model. The red line shown in Figure~\ref{fig:spec_extract} is the 1D extraction from the 2D drizzled model using the same aperture as the one used to extract the data.

We used the MCMC sampler from the Python package {\tt emcee} \citep{emcee+13} to obtain the posterior probability distributions on the three parameters. We used a flat prior on the redshift over $z\in[8.0,11.5]$, the interval to which the G141 grism data are sensitive for the LBG model. We used a Gaussian prior on $m160$, with center and standard deviation given by the isophotal magnitude and uncertainty in F160W. The isophotal magnitude is the correct normalization magnitude because we use the segmentation image to define the extent of the two-dimensional models. The ASTRODEEP catalog does not provide isophotal magnitudes, so we measured these independently using SExtractor in dual-image mode with the same parameter settings employed in Section~\ref{sec:contam}, finding $\mathrm{MAG\_ISO_{F160W}} = 25.63\pm0.06$. This is brighter than the total magnitude in the ASTRODEEP catalog reported in Table~\ref{tab:photometry}, which is due to a different estimation of the local background. ASTRODEEP performed an ICL subtraction on the F160W image, whereas we did not. While \obj{} is distant from the cluster center where the ICL contribution is the highest, there is still a small contribution near \obj{}, explaining why our magnitude is brighter than the ASTRODEEP magnitude. Because the isophototal magnitude is only used to normalize the MCMC models and is never compared to the ASTRODEEP photometry, this discrepancy does not affect our results. Finally, we adopted a half Gaussian prior on \ewlya{}, centered on zero with standard deviation of $15$ \AA{}, accepting only positive values because we are not sensitive to \lya{} absorption. We used the following log-likelihood function in the MCMC sampler:

\begin{eqnarray}
ln(\mathcal{L}) \sim -\chi^2/2 = -\frac{1}{2} \sum_i \frac{(f_i - m_i)^2}{\sigma_i^2}, \nonumber
\end{eqnarray}
where $f_i$, $m_i$, and $\sigma_i$ are the pixel values in the flux, model and uncertainty arrays, respectively. The uncertainty consists of the detector noise and the background noise (see Section~\ref{sec:extract}) added in quadrature. With this prescription, we found a reduced $\chi^2$ close to 1 for the best-fit parameters. We also experimented including contamination into the uncertainty array, but found that we over-fit the data by doing so. Therefore, we did not include contamination uncertainty in the uncertainty in the log-likelihood.

\begin{figure}[htb]
    \centering
    \includegraphics[width=\linewidth]{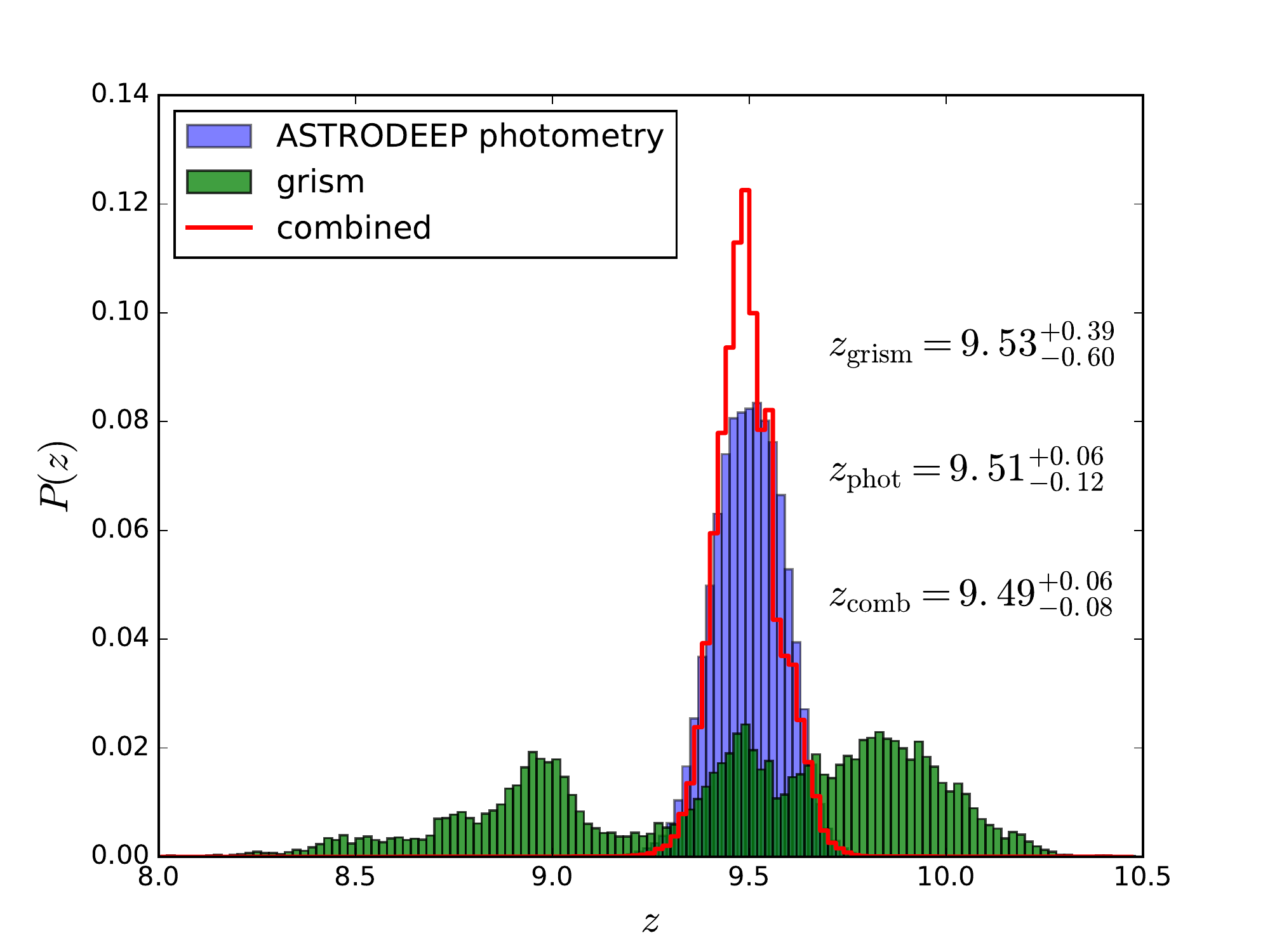}
        \includegraphics[width=\linewidth]{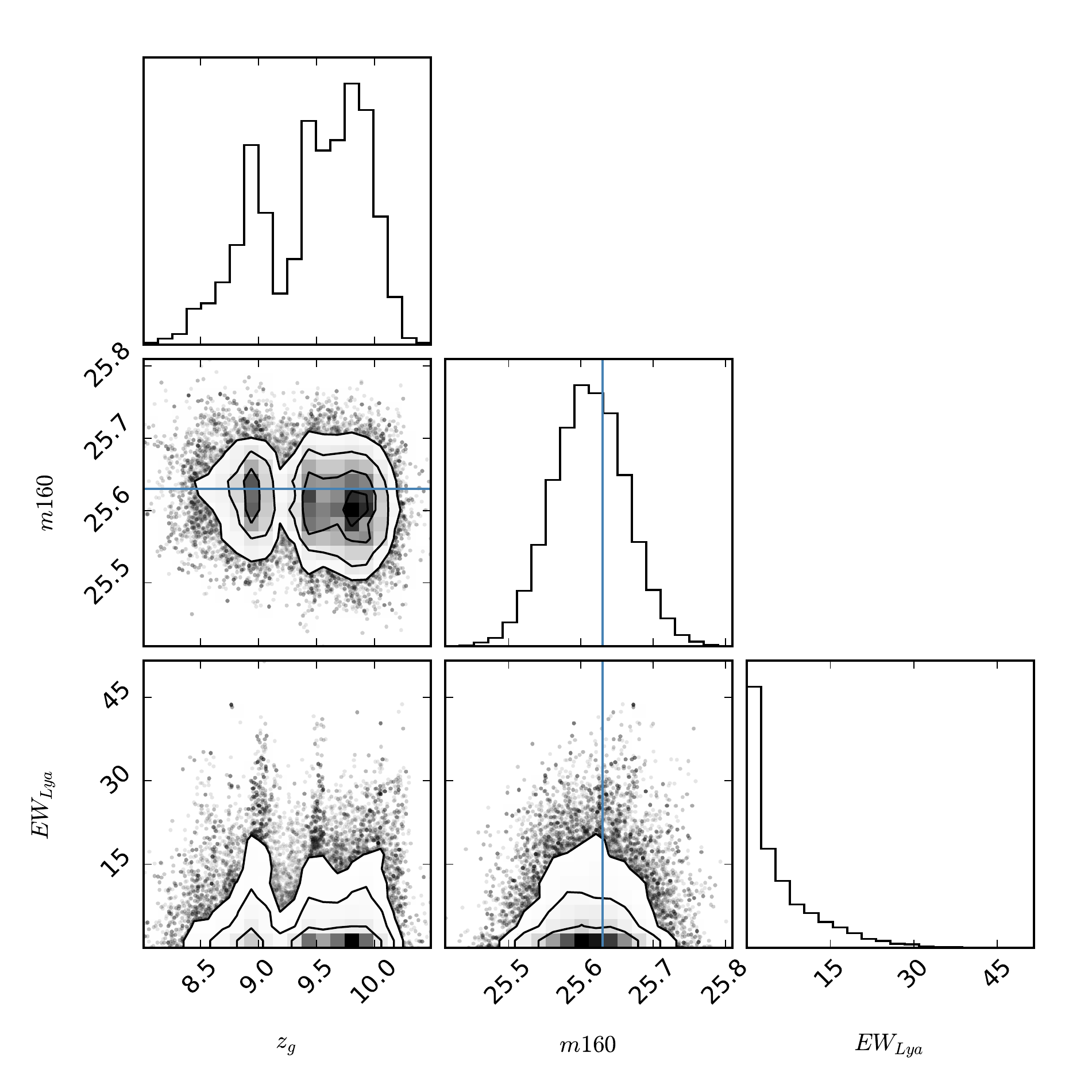}
    \caption{\textbf{Top}: Normalized redshift distributions inferred from photometry alone (blue histogram), from the grism alone (green histogram), and from multiplying the two distributions (red line). \textbf{Bottom}: Corner plot from MCMC fitting the grism data (excluding the contaminated Refsdal \refsdalpaI{}) to Lyman break galaxies models, where the three model parameters are the redshift $z_g$, the F160W normalization magnitude ($m160$), and the \lya{} equivalent width (\ewlya{}). The blue line indicates the value of MAG\_ISO measured in F160W, in good agreement with the F160W magnitude inferred from the grism data. }
 \label{fig:redshift}
 \end{figure}

The resulting corner plot for the 3 MCMC parameters is shown in Figure~\ref{fig:redshift}. The top left panel of the bottom sub-figure shows the marginalized likelihood for the redshift derived from the grism alone. This is also plotted on the top sub-figure for comparison with the photo-$z$. The best-fit redshift derived from the grism alone is \zgrismtext{} (68\% confidence), consistent with the photometric value of \zphottext{}. The grism redshift is much more uncertain than the photometric redshift, unsurprising given the level at which the continuum is detected in the grism. The combined likelihood is shown in red, with a best fit value of \zcombinedtext{} (68\% confidence). 

The other two MCMC parameters of the LBG model are $m160$ and \ewlya{}. The posterior for $m160$ is very similar to the prior, indicating that the grism spectra are not especially sensitive to this parameter. The fact that there is agreement with the photometric value serves as a sanity check. We derive an upper limit of $\mathrm{EW}_{Lya} < 6$ \AA{} ($1\sigma$) over the interval $8<z<11.5$, similar to what we found in the previous section using a different method ($\mathrm{EW}_{Lya} < 7$ \AA{}). This corresponds to a $1\sigma$ \lya{} flux limit of $1.5\times10^{-18}~\mathrm{erg\,s^{-1}cm^{-2}}$.
 
\section{Gravitational Lens Model}
\label{sec:lensmodel}

In order to constrain many of the intrinsic properties of \obj{}, a lens model of the galaxy cluster, \MJ{}, is required. Since the original gravitational lens models by \citet{Smith+09,Zitrin+09}, \MJ{} has been modeled extensively. As part of the version 1 (v1) HFF call for lensing models in 2013, before the HFF imaging took place, we created a strong and weak gravitational lens model of \MJ{} using the Strong and Weak Lensing United \cite[SWUnited;][]{Bradac+05,Bradac+09} code.  We have since modeled the cluster using constraints from the full-depth HFF imaging and \HST{} grism spectroscopy from GLASS and the SN Refsdal follow-up program \citep{Treu+16}. The magnification from the newer model presented by Finney et al. (in preparation) is $\mu_{\mathrm{best}}=$ \medmu{} (68\% confidence). This model will be referred to as the Bradac v4 model. Our magnification factor is in agreement with the median magnification from the six other publicly available post-HFF lens models\footnote{\url{http://www.stsci.edu/hst/campaigns/frontier-fields/Lensing-Models}}: $\mu=$ \medmuallmodels{}. 

 
\section{Discussion}
\label{sec:discussion}

\subsection{Physical properties of \obj{} }
\label{sec:phys}
The recovered physical properties of \obj{} are listed in Table~\ref{tab:phys}. These were obtained by using the combined grism$+$photometric redshift distribution as a prior in the SED fitting procedure outline in Section~\ref{sec:photoz}. As inferred by previous authors \citep{Zheng+12,Huang+16a,Zheng+17}, the galaxy is intrinsically faint and low mass due to its large, albeit uncertain, magnification. Using the Bradac v4 magnification factor obtained by Finney et al. (in preparation), $\mu_{\mathrm{best}}=$ \medmu{}, and the observed F160W magnitude, we calculated the rest-frame absolute UV magnitude to be $M_{\mathrm{UV}} - 2.5\, \mathrm{log}_{10}(\mu / \mu_{\mathrm{best}})=$ \medMUV{} mag, or $<0.15L_{\star} (1\sigma)$, adopting $L_{\star}$\footnote{The characteristic luminosity has not been reliably measured at higher redshift due to small sample sizes at $z=7.9$ from \cite{Bouwens+15}}. The factor $\mu / \mu_{\mathrm{best}}$ is introduced to parameterize our uncertainty on the magnification.

The most striking property of \obj{} is its relatively large age of \medAGE{} $\mathrm{Myr}$. The inferred age is primarily driven by the red color in \spitzer{}/IRAC: $\mathrm{CH1} - \mathrm{CH2} > 0.73$ $(3\sigma)$, which was interpreted as a Balmer/4000 \AA{} break by the SED fitting code (Figure~\ref{fig:SED}). We note that the inferred age in general depends on the assumed SFH. We used an exponentially declining SFH to derive the age of \medAGE{} $\mathrm{Myr}$. Adopting an exponentially rising SFH instead, we infer an age consistent within the statistical uncertainties of the age derived using exponentially declining SFH. The age of the galaxy, given the peak combined redshift, would indicate an onset of star-formation of only $\sim200$ Myr after the Big Bang. By comparison, O16 spectroscopically confirmed a galaxy at higher redshift ($z=11.09$), yet the age of $<110$ Myr ($1\sigma$) implies an onset of star formation at $\gtrsim300$ Myr after the Big Bang, later than what is inferred for \obj{}. 

From the theoretical side, the inferred age of \medAGE{} $\mathrm{Myr}$ at $z=9.5$ is highly improbable. Using the luminosity function model from \citet{Trenti+15}, we computed the halo assembly time for a $log_{10}(M_h/M_{\odot})=10.5$ halo\footnote{The assembly time is only weakly dependent on the halo mass} at $z=9.5$, finding a median time of $65~\mathrm{Myr}$, and an upper limit of $<155~\mathrm{Myr}$ (99\% confidence). The fact that the inferred age of \obj{} is much larger than the upper limit on the assembly time is therefore surprising. If the age estimate is correct, it would require star formation well before the halo assembly time. 

\subsection{Interpretation of rest-frame optical observations}
The age estimate of \medAGE{} $\mathrm{Myr}$ is strongly dependent on the photometric redshift estimate, which is driven by the interpretation of the IRAC color as a continuum break. Recent observations of $z\sim7$ LBGs have revealed that this population on average has strong rest-frame optical  emission lines that are capable of boosting IRAC CH1 and/or CH2  fluxes by up to $\sim1$ mag \citep[e.g.][]{Schaerer+09,Shim+11,Labbe+13,deBarros+14,Smit+14,Castellano+17}. For example, \cite{Smit+14} showed that $z\sim7$ LBGs have average \oiii{}$+H\beta$ equivalent widths of $>640$ \AA{}. We investigated whether \obj{} could possess similarly large emission line strengths that could explain the IRAC color of $\mathrm{CH1} - \mathrm{CH2} > 0.73$ $(3\sigma)$. This would clearly change the physical interpretation of the galaxy, particularly the age. 

First, we estimated the photo-$z$ without the IRAC CH1 and CH2 fluxes, eliminating the rest-frame optical constraints on the SED, which could bias the redshift inference. The 95\% confidence interval of the resulting redshift likelihood function was $[8.93-9.77]$, about a factor of 2 broader than when including the IRAC data. This is consistent with the photometric redshift obtained by \cite{Kawamata+16}, $z=9.2\pm1.0$, who used the full depth HFF data but no IRAC or Ks-band data. We then multiplied this distribution with the grism likelihood distribution obtained in Section~\ref{sec:mcmc}. The 95\% confidence interval of the combined likelihood function was $[8.95-9.77]$, very similar to the range from the photometry alone. Over this interval, \oii{} always falls in CH1, $H\beta$ falls in CH2 for $z\lesssim9.4$, and \oiii{} falls in CH2 for $z\lesssim9.2$.

We explored adding \oii{}$\lambda3727$, [\emissionline{O}{iii}]$\lambda\lambda4959,5007$, H$\beta$, the remaining Balmer lines, and common metal lines \citep[see Table 1 of][]{Anders+03} in emission onto a redshifted $10~\mathrm{Myr}$ template spectrum (with no emission lines originally added) generated from the \textit{BC03} models assuming $Z=0.2 Z_{\odot}$, $Z=0.02 Z_{\odot}$, and $Z=Z_{\odot}$ to see if we could reproduce the IRAC color. We added in the $H\beta$ emission line, varying the rest-frame EW from $0-10000$ \AA{}, and we calculated the rest-frame EWs of the other lines using the fixed flux ratios determined by \citet{Anders+03}, given the template metallicity. The continuum flux densities in CH1 and CH2 used to convert between EW and line flux were determined by integrating the templates over the respective bandpasses. We explored the redshift range given by the 95\% confidence interval on the combined redshift derived above $z\in[8.95-9.77]$ in steps of $\Delta z = 0.05$. 

At $z>9.3$, the IRAC color cannot be recovered with $\mathrm{EW}_{H\beta} < 10000$ \AA{} (equivalently:  $\mathrm{EW}_{[OIII]+H\beta} \lesssim 50000$ \AA{}) for any of the input metallicities, as expected because the strongest emission line that we included, the \oiii{} pair, falls redward of CH2. Therefore, strong rest-frame optical lines are unlikely to cause the red IRAC color if the redshift is $z>9.3$. At $9.05\leq z \leq 9.3$, we find that $\mathrm{EW}_{H\beta} \sim 800-1100$ \AA{} ($\mathrm{EW}_{[OIII]+H\beta} \sim 4300-5800$ \AA{}) are required to reproduce the IRAC color, larger than the equivalent widths observed by \citet{Smit+14}. We note that these rest-frame equivalent widths are for the $Z=0.02 Z_{\odot}$ case, with higher metallicities requiring much larger equivalent widths to reproduce the IRAC color.

However, in the range $8.95\leq z<9.05$, we are able to reproduce the IRAC color with precedented values of $\mathrm{EW}_{H\beta}$.The equivalent widths required to do so are $\mathrm{EW}_{H\beta} \sim 425-550$ \AA{} ($\mathrm{EW}_{[OIII]+H\beta} \sim2200-2900$ \AA{}) for $Z=0.02Z_{\odot}$ or $\mathrm{EW}_{H\beta} \sim 350-375$ \AA{} ($\mathrm{EW}_{[OIII]+H\beta} \sim2500-2800$ \AA{}) for $Z=0.2Z_{\odot}$. These equivalent widths are consistent with the sample from \citet{Smit+14}, where they measured $\mathrm{EW}_{[OIII]+H\beta}$ from \HST{}-selected $z\sim7$ galaxies in cluster fields. They are also consistent with $\mathrm{EW}_{[OIII]+H\beta}$ measured for the Green Peas \citep{Henry+15} and from metal-poor local dwarf galaxies specifically selected to show \emissionline{He}{ii} \citep{Senchyna+17}. We note, however, that there is only a 5\% chance that the redshift is $z<9.05$, based on the joint grism$+$photometric redshift distribution once IRAC constraints are dropped. While one of the peaks of the grism redshift distribution occurs around $z\sim9$, the \HST{} photometry (even after neglecting the IRAC constraints) provide the tightest constraints against lower redshifts. This exercise demonstrates that the \HST{} photometry, \spitzer{}/IRAC photometry and grism spectra are all consistent with a $10~\mathrm{Myr}$, sub-solar metallicity galaxy with strong rest-frame optical nebular emission lines, but only in the low-redshift tail of the redshift distribution. 

The reason our SED-fitting process was unable to fit nebular emission lines with such large equivalent widths is due to the manner in which they were added to the \textit{BC03} templates. We used the relation from \citet{Leitherer+95} to scale the Lyman continuum flux predicted by the \textit{BC03} templates to the $H\alpha$ flux. We then used the flux ratios from \citet{Anders+03} given the template metallicity to get the fluxes of the other nebular emission lines. For a 10 Myr old, $Z=0.02 Z_{\odot}$ template, the maximum  rest-frame $\mathrm{EW}_{H\beta} \sim 100$ \AA{}. This is $\sim250$ \AA{} lower than the minimum rest-frame $\mathrm{EW}_{H\beta}$ necessary to produce the IRAC color. At higher metallicities, the emission line strengths predicted from the models are even lower. While we used $Z=0.2 Z_{\odot}$ templates when performing SED fitting, a lower metallicity template similarly would not have been able to fit the IRAC photometry with strong nebular emission lines. Because larger EWs are observed in practice at $z\sim7$ \cite[e.g.][]{Smit+14}, the prescription we use for adding nebular emission may be inappropriate here. Rest-frame optical spectroscopy with the \textit{James Webb} Space Telescope of $z\gtrsim7$ galaxies will provide direct measurements of the nebular line strengths. These measurements will be able to break the degeneracy between the Balmer/$4000$ \AA{} break and strong rest-frame optical emission lines for \obj{}. 


\begin{deluxetable}{lcl}
\tablecaption{Properties of \obj{} }
\tablecolumns{3}
\tablewidth{\linewidth}

\startdata
\hline
$\mathrm{R.A.}$ & ($^{\circ}$) &  177.38997   \\
$\mathrm{Dec.}$ & ($^{\circ}$) &  22.412722   \\
$z_{\mathrm{grism+photometry}}$ & & \zcombined{}  \\
$\mu_{\mathrm{best}} (z=$\zcombined{}) & & \medmu{} \\
$M_{\mathrm{UV}} - 2.5\, \mathrm{log}_{10}(\mu / \mu_{\mathrm{best}})$ & (mag) & \medMUV{} \\ 
$M_{\star} \times \mu / \mu_{\mathrm{best}}$ & ($10^{8} M_{\odot}$) & \medsmass{} \\
$L_{\mathrm{UV}} \times \mu / \mu_{\mathrm{best}} $ & ($L_{\star}$\tablenotemark{a}) & \medLUV{} \\
$SFR \times \mu / \mu_{\mathrm{best}}$ & ($M_{\odot} / \mathrm{yr} $) & \medSFR{} \\
Age & ($\mathrm{Myr}$) & \medAGE{} \\
$E(B-V)$ & mag & $ < 0.01$ \\
\ewlya{} ($3\sigma$) & \AA{} & $<21$
\enddata
\tablecomments{The absolute rest-frame UV magnitude is determined using the F160W magnitude to calculate the underlying continuum flux density. The magnification used to correct the absolute UV magnitude, stellar mass and star formation rate are from the Bradac v4 model. }
\tablenotetext{$a$}{Characteristic luminosity at $z=7.9$ from \citet{Bouwens+15}. $L_{\star}$ has not been reliably measured at higher redshift due to the small sample size. }
\label{tab:phys}
\end{deluxetable}


\section{Summary}
\label{sec:summary}

We presented deep grism-spectroscopic observations of \obj{}, one of the highest redshift galaxies known. We tentatively detected the rest-frame UV continuum in the combined \HST{} G141 data, although contamination residuals near the trace of \obj{} make this detection somewhat uncertain. We used an MCMC fitter to constrain a simple LBG model with three parameters, ($z$, F160W mag, and rest-frame \lya{} EW). We found that the redshift and F160W magnitude inferred by the grism data are consistent with our photometric measurements, as well as those from the literature using the same quality photometry. The grism data provide the most constraining upper limit on \lya{} emission for this galaxy to date, ruling out \lya{} in emission above an equivalent width of $21$ \AA{} ($3\sigma$).

The grism data support the interpretation of the red \spitzer{}/IRAC $\mathrm{CH1} - \mathrm{CH2}$ color as a Balmer/4000 \AA{} break, as well as the strong rest-frame optical emission lines boosting the CH2 flux. The strong emission line scenario is less preferred by the \HST{} photometry but cannot be completely ruled out. The preferred interpretation implies that the galaxy fosters a stellar population that is at least $\sim300~ \mathrm{Myr}$ old already at $z=$\zcombined{}, the combined grism$+$photometric redshift. If this interpretation of the IRAC color is correct, then this implies that \obj{} formed around $200~\mathrm{Myr}$ after the Big Bang, making it one of the earliest sites for star-formation known to date.

\vspace*{0.5cm}

\acknowledgments{A.H. and this work were supported by NASA (National Aeronautics and Space Administration) Headquarters under the NASA Earth and Space Science Fellowship Program, Grant ASTRO14F-0007. Support for the Grism Lens-Amplified Survey from Space (GLASS) (HST-GO-13459) was provided by NASA through a grant from the Space Telescope Science Institute (STScI). Support for this work was also provided by NASA through an award issued by JPL/Caltech (for SURFS UP project) and by {\HST}/STScI  {\HST}-AR-13235, {\HST}-AR-14280, and {\HST}-GO-13177. This work utilizes gravitational lensing models produced by PIs Brada\v{c}, Richard, Natarajan \& Kneib, Sharon, Williams, Keeton, Bernstein and Diego, and Oguri. This lens modeling was partially funded by the HST Frontier Fields program conducted by STScI. STScI is operated by the Association of Universities for Research in Astronomy, Inc. under NASA contract NAS 5-26555. The lens models were obtained from the Mikulski Archive for Space Telescopes (MAST).}

\appendix

\section*{Testing the validity of the continuum detection in G141}
\label{sec:appendixA}

\renewcommand\thefigure{A\arabic{figure}} 
\setcounter{figure}{0}    

In this appendix we show several tests for determining the validity of the continuum detection in the presence of contamination. 
 
\begin{figure}[htb]
    \centering
   \includegraphics[width=\linewidth]{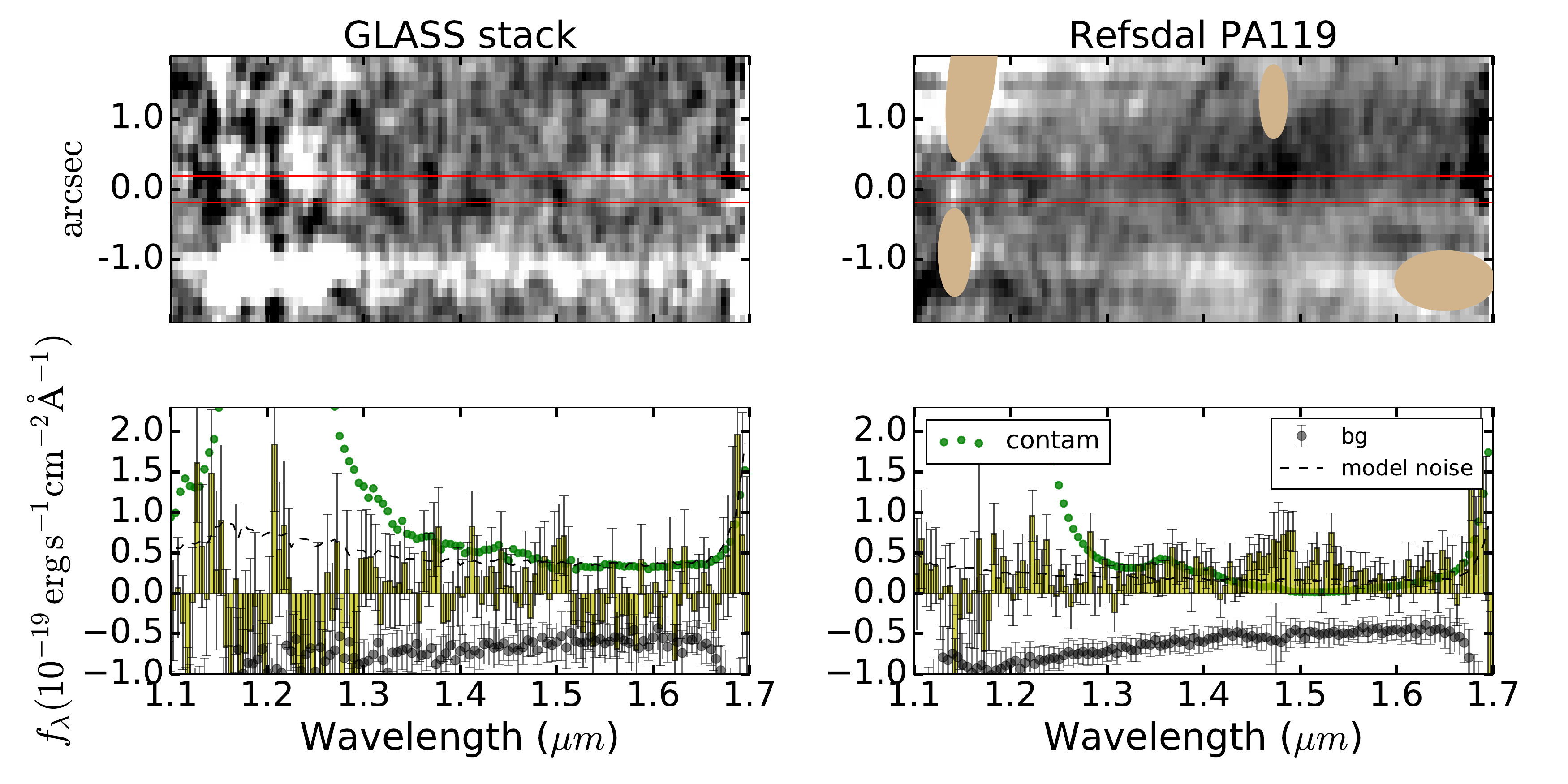}
    \caption{\textbf{Top}: Contamination-subtracted two-dimensional grism spectra from the two GLASS (left) and single Refsdal \refsdalpaII{} (right) position angles. As in Figure~\ref{fig:spec_extract}, black represents positive signal and the 2D spectra were smoothed with a 2D Gaussian kernel with $\mathrm{FWHM}=2$ pixels in both dimensions. \textbf{Bottom}: One-dimensional extractions (yellow histogram) of the two-dimensional spectra above. The bulk of the signal comes from the deeper Refsdal data, as expected for a real signal. The strong negative residual at $\sim-1.2$ arcsec in the GLASS stack comes almost entirely from the GLASS \gpaI{}. This residual is down-weighted in Figure~\ref{fig:spec_extract} when making the stack. While in principle the signal should taper at $\lambda \geq 1.65~\mu m$ due to the declining throughput of G141, contamination is severe in both cases at these wavelengths, making this test uninformative.}   
 \label{fig:glassvsrefsdal2}
 \end{figure}

If the continuum detection in the stack of all three orients shown in Figure~\ref{fig:spec_extract} is real, it should come primarily from the deeper Refsdal \refsdalpaII{} data. We show the spectra from combining the two GLASS orients as well as the individual Refsdal \refsdalpaII{} data in Figure~\ref{fig:glassvsrefsdal2}. The stacked GLASS spectra are consistent with the noise, whereas the Refsdal \refsdalpaII{} data are consistently above the noise. Thus the majority of the signal in the stack comes from the Refsdal \refsdalpaII{} data. We verified that the signal does not arise from a single exposure or visit within the Refsdal \refsdalpaII{} data. When divided in half, the Refsdal \refsdalpaII{} data show a similar signal in both data sub-sets.  

The spectra in Figure~\ref{fig:glassvsrefsdal2} extend out to $1.7~\mu m$, the edge of the G141 grism. At $\lambda \geq 1.65~\mu m$, the throughput of G141 drops quickly, which should in principle be reflected by the spectrum of a real object. In our case, contamination rises in both the GLASS and Refsdal \refsdalpaII{} at these wavelengths. As a result, we cannot reliably use this test to determine the validity of our spectrum.

The continuum magnitude derived in the grism is somewhat brighter than the expectation from the photometry. For example, if we integrate the grism flux density from the 3 good position angles within a $0\farcs38$ spatial aperture over the F160W bandpass, which is entirely contained in G141, we get a magnitude of F160W (grism) $= 25.5\pm0.1$, compared to a photometric magnitude of F160W (imaging) $=26.1\pm0.1$ using a spatial aperture of the same size. We note that this photometric magnitude is fainter than the magnitude reported in Table~\ref{tab:photometry} because of the smaller aperture used in this test. The photometric magnitude is measured by creating a flat spectrum that is normalized to F160W $=\mathrm{MAG\_ISO}$, and then convolving this spectrum with the segmentation image of \obj{}, drizzling it to a grid with the same dimensions and position angles as the data and extracting the magnitude within the same aperture as the grism magnitude. We find a difference in grism and photometric magnitude of F160W (grism) $-$ F160W (imaging) $=-0.6\pm0.1$ mag. 

\begin{figure}[htb]
    \centering
    \includegraphics[width=0.75\linewidth]{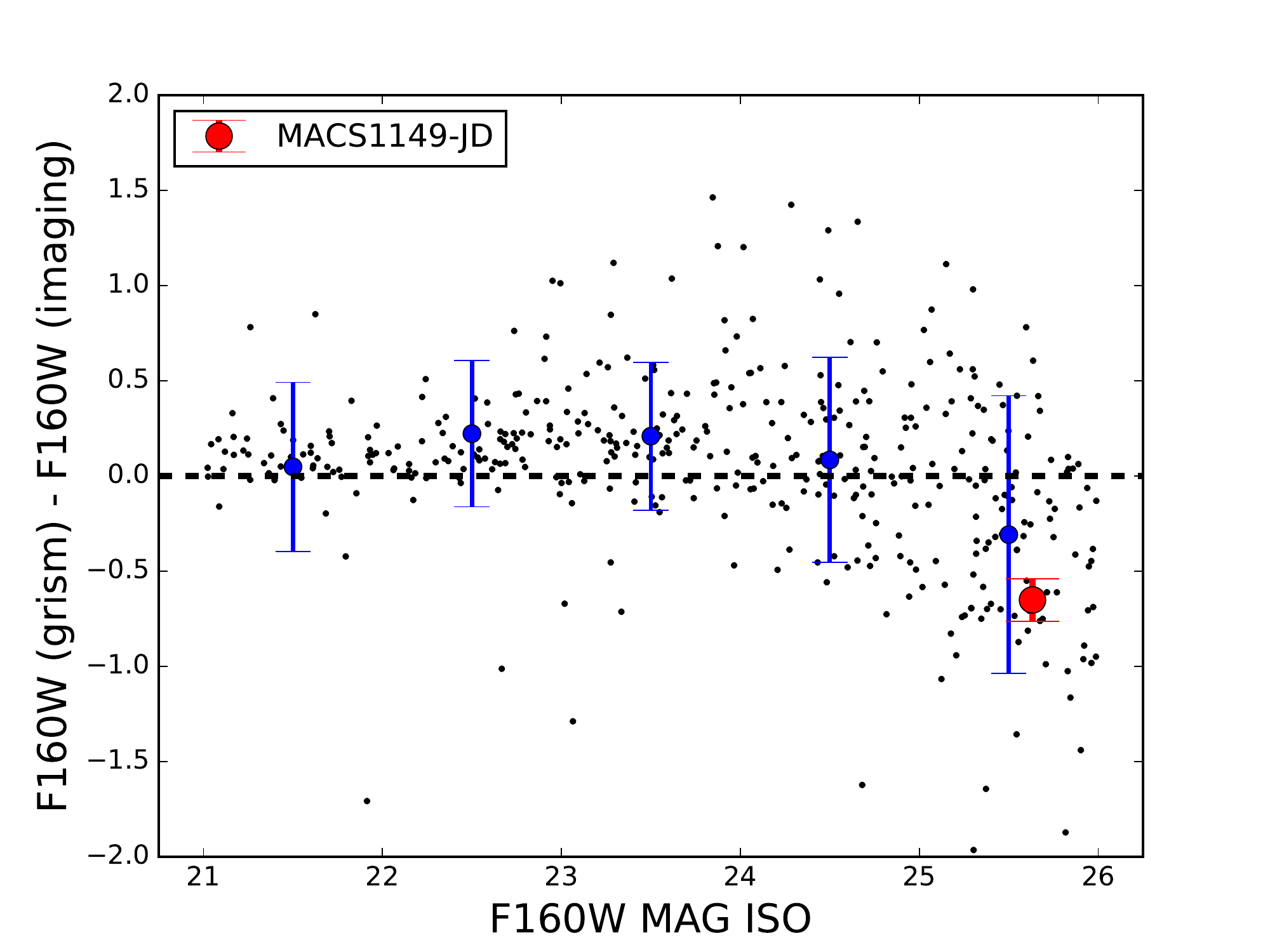}
    \caption{Comparison of magnitudes calculated from the grism and from the photometry over the \HST{} F160W bandpass for 425 individual objects (black points) in the grism field of view. The blue points and error bars represent the mean and standard deviation of the points in equally spaced bins that are 1 magnitude wide. The magnitude offset for \obj{} is shown in red, consistent with the offsets observed for similarly faint objects. }
 \label{fig:compare_mags}
 \end{figure}

To test the relative flux calibration between the grism and photometry in general, we carried out the same comparison for all objects with $21 < \mathrm{MAG\_ISO} < 26$ in the segmentation map that had full coverage in the 3 position angles used to extract \obj{} (both GLASS and Refsdal \refsdalpaII{}) that had a match within $0\farcs1$ in the photometric catalog. 533 objects obeyed these criteria. However, a negative flux was recovered for 108 of these, particularly at the faint end, so we only used the remaining 425 objects in the comparison. Figure~\ref{fig:compare_mags} shows the comparison of the grism and photometric magnitudes of these 425 objects as a function of catalog $\mathrm{MAG\_ISO}$. The grism magnitudes tend to recover the photometric magnitude accurately up until $\mathrm{MAG\_ISO} \simeq=25$, although with large scatter. At $\mathrm{MAG\_ISO} >25$, the grism flux on average predicts a higher F160W magnitude than inferred from the photometry, and the scatter increases to $\sim0.7$ mag, about 7 times larger than the statistical uncertainty on the magnitude offset for \obj{}. The magnitude offset observed for \obj{} is consistent with the general offsets seen at similarly faint magnitudes.

\begin{figure}[htb]
    \centering
   \includegraphics[width=0.95\linewidth]{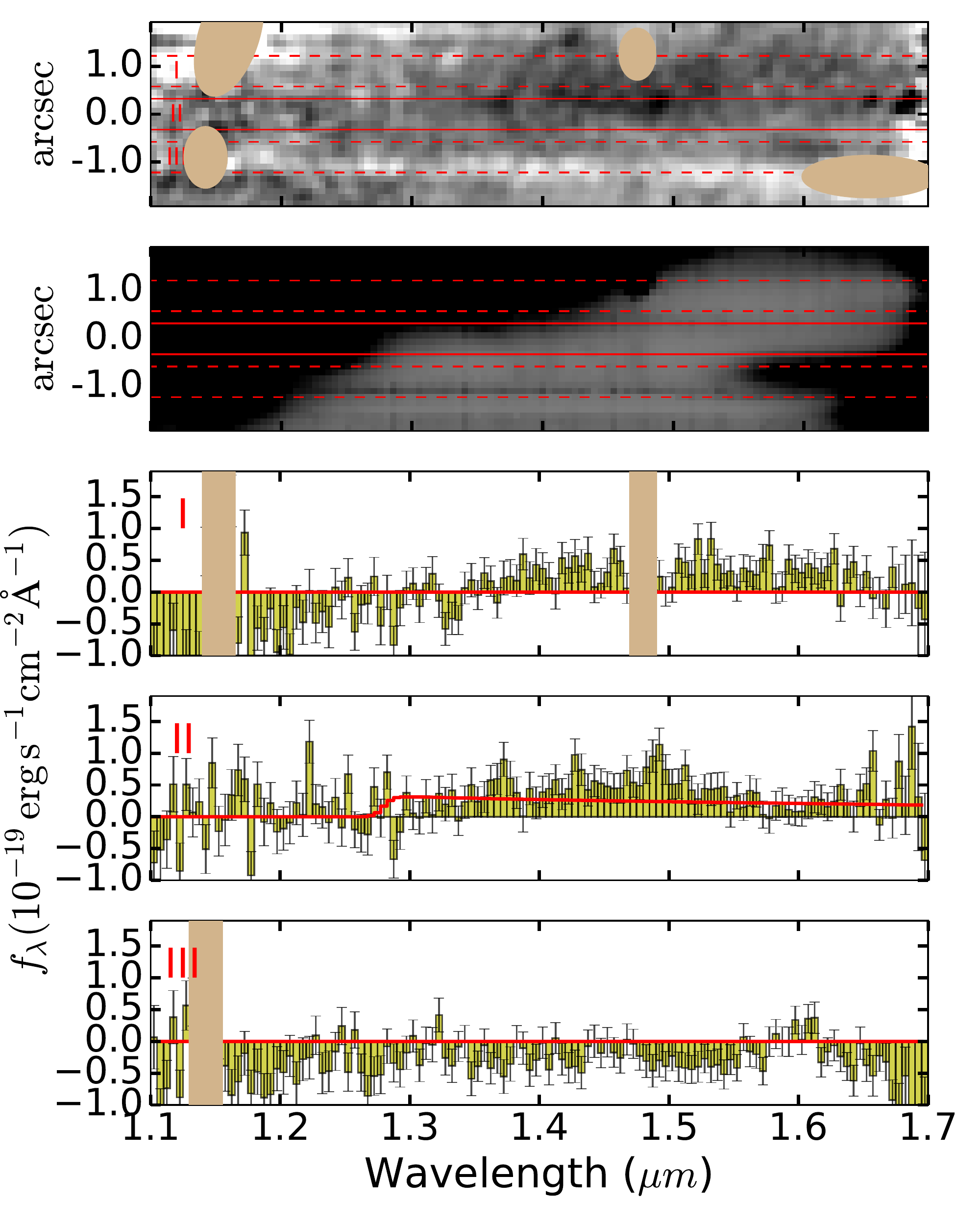}
    \caption{Drizzled two-dimensional spectrum (top), contamination model (second row), and one-dimensional extractions at three different vertical extraction apertures from the combination of the two GLASS P.A.s and the Refsdal \refsdalpaII{} data. The top spectrum was smoothed with a 2D Gaussian kernel with $\mathrm{FWHM}=2$ pixels in both dimensions. The spatial apertures (I, II and III) are chosen so that the flux from \obj{} predicted from \HST{} F160W falls within the central (II) aperture only. Contamination residuals are strong everywhere except the central aperture where the trace from \obj{} is expected. }
 \label{fig:offsets}
 \end{figure}

We also extracted one-dimensional spectra at vertically offset positions from the trace of \obj{}. In Figure~\ref{fig:offsets}, we show the cleaned two-dimensional spectrum (as in Figure~\ref{fig:spec_extract}), the contamination model for this combination of position angles, as well as the one-dimensional spectrum extracted in three different spatial apertures, labeled I$-$III on the two-dimensional spectra. The red line in each of the three bottom panels shows the expected signal from the model, which should be very close to zero in panels I and III. However, positive residuals in panel I and negative residuals in panel III cause a significant deviation from zero. The strong negative signal in panel III is almost entirely accounted for by the residual from a single contaminant in the GLASS \gpaI{} spectrum, which is visible in the contamination panel. While panels I and III are contaminated, it is reassuring that panel II, in which the trace of \obj{} is expected, is relatively clean of contamination. While we expect that the signal present in this panel likely arises from \obj{}, the off-trace extractions indicate that our contamination model could be problematic, even in panel II.

\begin{figure}[htb]
    \centering
    \includegraphics[width=\linewidth]{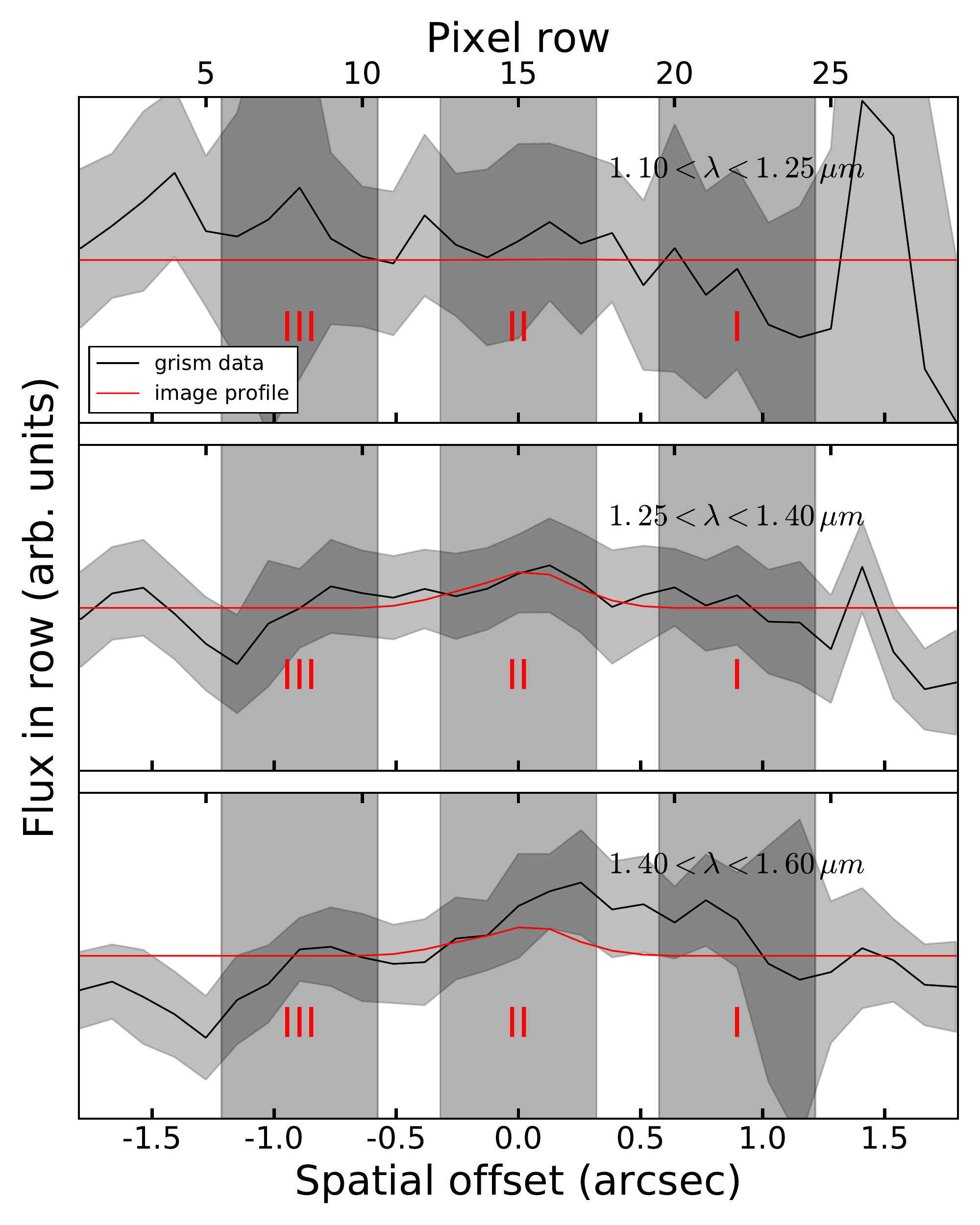} 
    \caption{The spatial profile extracted from the Refsdal \refsdalpaII{} spectrum of \obj{}. The three panels show the spatial extraction in three separate wavelength intervals: \textbf{top}: $1.10<\lambda<1.25~\mu m$, \textbf{middle}: $1.25<\lambda<1.40~\mu m$, \textbf{bottom}: $1.40<\lambda<1.60~\mu m$. In all three panels, the black line and dark gray region represent the mean and standard deviation, respectively, of the flux in each spatial row of the two-dimensional spectrum. The red line indicates the spatial profile extracted from a LBG model fixed to the best-fit grism redshift, \zgrismpeaktext{}, whose morphology and magnitude are determined by the \HST{} F160W image. The three spatial regions from Figure~\ref{fig:offsets} are shown on each panel with shaded vertical regions. Reassuringly, the data show a peak in flux in the central region (region II) for the two redder wavelength bins, but not in the bluer bin, consistent with the expectation from the model. Residual contamination is likely responsible for the excess flux in the bottom panel at positive spatial offsets. }
 \label{fig:refsdal2_spatial}
 \end{figure}

To assess the impact of contamination residuals on the interpretation of the spectrum, we also looked at the spatial profile of the signal. Figure~\ref{fig:refsdal2_spatial} shows the collapsed spatial profile of the source as well as the data for the single Refsdal \refsdalpaII{} in three different wavelength intervals. The single PA was chosen so that the spatial profile of the source was well-defined. In the two wavelength intervals (bottom two panels) where flux is expected from the model, the data exhibit flux in the central spatial bin, consistent with the model. In the other wavelength interval (top panel) where no flux is expected from the model, $1.10<\lambda<1.25~\mu m$, the data are consistent with no flux. Excess flux at positive spatial offsets in the reddest wavelength interval, $1.40<\lambda<1.60~\mu m$, is likely due to residual contamination. A second order spectrum of a bright object lands in this part of the \obj{} spectrum. The fit to the first order of this object is poor due to the presence of unmodeled contamination along its own trace. The unmodeled contamination likely arises from objects outside of the field of the segmentation image.
 
 
\end{document}